\documentclass[10pt,journal,compsoc]{IEEEtran}
\usepackage{url}
\usepackage{color}
\usepackage{graphicx}
\usepackage{caption2}
\usepackage[linesnumbered,ruled]{algorithm2e}
\usepackage{multirow}
\usepackage{amsmath}
\usepackage{amsfonts}
\urldef{\mailsa}\path|{wbzhao, dbzhao}@hit.edu.cn|
\urldef{\mailsb}\path|xmliu.hit@gmail.com|
\urldef{\mailsc}\path|shiqwang@cityu.edu.hk|

\def\W{\mathbf{W}}

\def\n{{\mathbf n}}
\begin{document}

\title{NormalNet: Learning-based Normal Filtering for Mesh Denoising}

\author{Wenbo Zhao,
Xianming Liu,~\IEEEmembership{Member,~IEEE},
Yongsen Zhao,
Xiaopeng Fan,~\IEEEmembership{Senior Member,~IEEE}
Debin Zhao~\IEEEmembership{Member,~IEEE},

\thanks{
W. Zhao, X. Liu, Y. Zhao, X. Fan and D. Zhao are with the School of Computer Science and Technology, Harbin Institute of Technology, Harbin, 150001, China. e-mail: \{wbzhao, csxm, 18S103199, fxp, dbzhao\}@hit.edu.cn.

}
}


\IEEEtitleabstractindextext{

\begin{abstract}
Mesh denoising is a critical technology in geometry processing that aims to recover high-fidelity 3D mesh models of objects from their noise-corrupted versions. In this work, we propose a learning-based normal filtering scheme for mesh denoising called \emph{NormalNet}, which maps the guided normal filtering (GNF) into a deep network. The scheme follows the iterative framework of filtering-based mesh denoising. During each iteration, first, the voxelization strategy is applied on each face in a mesh to transform the irregular local structure into the regular volumetric representation, therefore, both the structure and face normal information are preserved and the convolution operations in CNN(Convolutional Neural Network) can be easily performed. Second, instead of the guidance normal generation and the guided filtering in GNF, a deep CNN is designed, which takes the volumetric representation as input, and outputs the learned filtered normals. At last, the vertex positions are updated according to the filtered normals. Specifically, the iterative training framework is proposed, in which the generation of training data and the network training are alternately performed, whereas the ground truth normals are taken as the guidance normals in GNF to get the target normals. Compared to state-of-the-art works, \emph{NormalNet} can effectively remove noise while preserving the original features and avoiding pseudo-features.
\end{abstract}

}

\maketitle

\IEEEdisplaynontitleabstractindextext

\IEEEpeerreviewmaketitle

\IEEEraisesectionheading{\section{Introduction}\label{sec:introduction}}

\IEEEPARstart{R}{ecently}, a demand for high-fidelity 3D mesh models of real objects has appeared in many domains, such as computer graphics, geometric modelling, computer-aided design and the movie industry.  However, due to the accuracy limitations of scanning devices, raw mesh models are inevitably contaminated by noise, leading to corrupted features that profoundly affect the subsequent applications of meshes. Hence, mesh denoising has become an active research topic in the area of geometry processing.

Mesh denoising is an ill-posed inverse problem. The nature of mesh denoising is to smooth a noisy surface, while concurrently preserving the real object features without introducing unnatural geometric distortions. Mesh denoising is a challenging task, especially for cases with large and dense meshes and with high noise levels. The key to the success of mesh denoising is to differentiate the actual geometry features, such as the localized curvature changes and small-scale details, and the noise generated by scanners. The literature contains rich work on mesh denoising, including filtering-based~\cite{yagou2002mesh,wei2015bi,zhang2015guided,W11,W12,Li2017,GGNF}, feature-extraction-based~\cite{lu2016robust,wei2016tensor}, optimization-based~\cite{he2013mesh,wang2014decoupling}, and similarity-based~\cite{yoshizawa2006smoothing,rosman2013patch,digne2012similarity}. Among these methods, the guided-normal based scheme has become popular in recent years~\cite{zhang2015guided,W11,W12,Li2017,GGNF}, which follows the iterative framework of filtering-based mesh denoising. During each iteration, the guidance normals are derived and used in filtering first. Then the vertex positions are updated according to the filtered normals. This approach performs mesh denoising either by building guidance normals with manually designed methods~\cite{zhang2015guided,Li2017,GGNF}, or by introducing additional information to improve the performance of guidance normals~\cite{W11,W12}. The schemes in ~\cite{zhang2015guided,W11,W12} perform well on synthetic meshes with simple structures. However, the methods of generating guidance normals in~\cite{zhang2015guided,W11,W12} are based on finding consistent patches with fixed shapes, therefore cannot handle complex structures well such as narrow edges and corners. To overcome this problem, Li \textit{et al.}~\cite{Li2017} propose to generate the guidance normals by the corner-aware and edge-aware neighbourhood. Zhao \textit{et al.}~\cite{GGNF} employ the graph-cut to generate piece-wise smooth patches and build guidance normals on them. These schemes perform well on synthetic meshes with complex features. However, the main idea of these schemes is finding consistent patches according to the face normal difference, and the structure information has not been fully utilized. For scanned meshes, which contain manifold kinds of noise and more complex shapes, such as serrated noise (Fig.~\ref{r5}), stair-stepping noise (Fig.~\ref{r8}) and irregular edges (Fig.~\ref{r4}), the face normal difference of noisy faces is so large that it is difficult for~\cite{Li2017,GGNF} to distinguish noise and features, resulting in either introducing pseudo-features or over-smooth. As shown in the experimental comparisons, even the state-of-the-art schemes~\cite{zhang2015guided,GGNF} cannot handle these cases well.

In the counterpart 2D image denoising, deep-learning-based strategies, such as~\cite{Chen2015Trainable,Zhang2017FFDNet,Zhang2017Beyond}, have been widely applied and achieved great success. However, with respect to mesh denoising, to the best of our knowledge, no studies follow this line of research. One main reason preventing the usage of convolutional neural network(CNN) in mesh denoising is that, in contrast to the regular grid structure of 2D images, meshes have irregular structures. Therefore, it is not straightforward to apply the regular 3D convolutional operations in CNNs to a mesh. Another reason may be the difficulty of selecting an efficient denoising strategy for CNN to mimic.

\begin{figure*}
\centering
\includegraphics[width=0.95\linewidth]{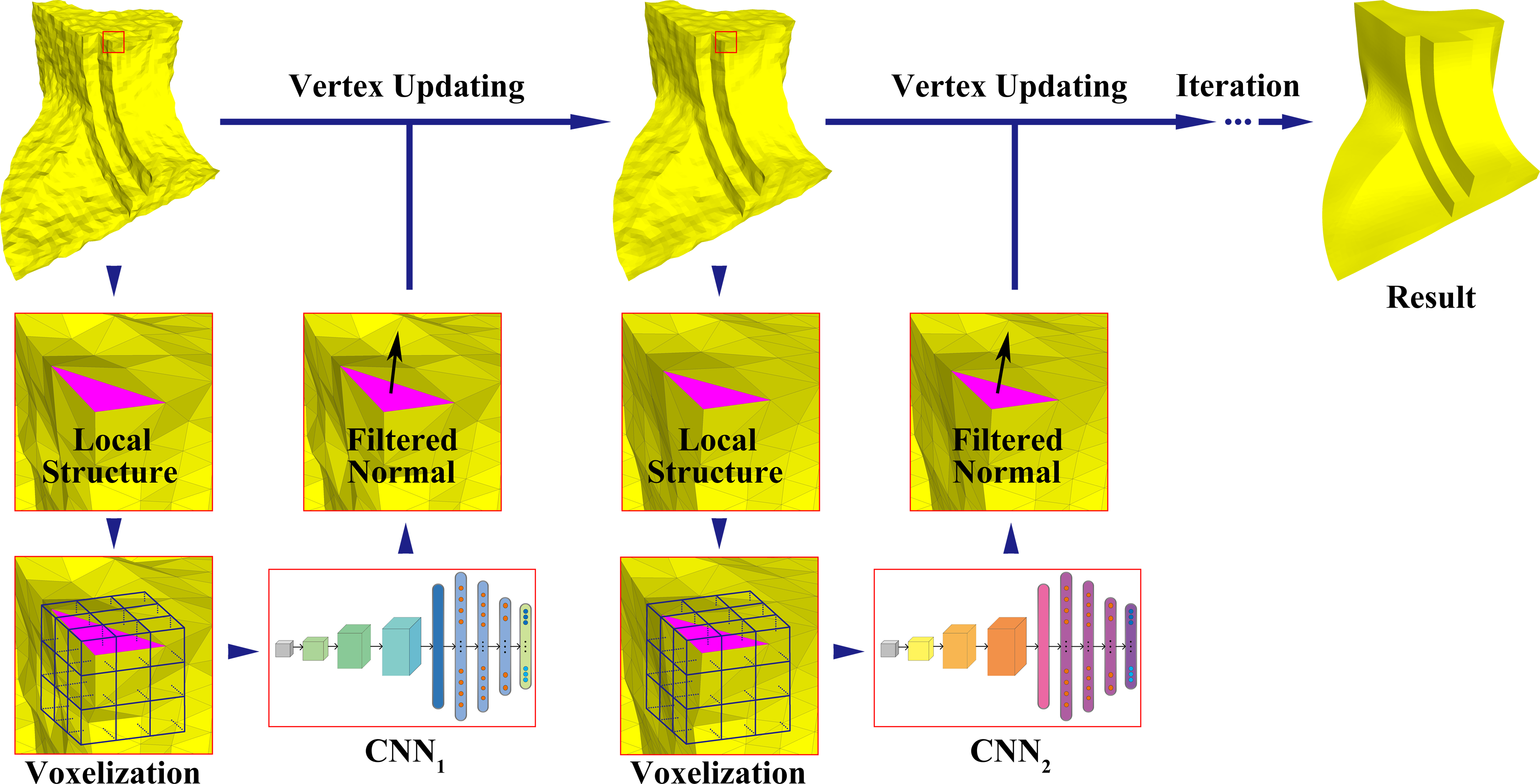}
\caption{The framework of \emph{NormalNet}. Modules in all iterations share the same workflow: for a face, the irregular local 3D structure is converted via the voxelization strategy into the regular volumetric representation, which is then input into CNN to get the filtered normal. Finally, the vertex positions are updated to obtain the denoised mesh.}
\label{fig_overview}
\end{figure*}
In this work, we propose a learning-based normal filtering scheme for mesh denoising called \textit{NormalNet}, which maps the guided normal filtering (GNF)~\cite{zhang2015guided} into a deep network. In particular, \textit{NormalNet} follows the iterative framework of GNF, as shown in Fig.~\ref{fig_overview}. During each iteration, to overcome the difficulty in using CNNs on meshes and exploit both the structure and face normal information, first, the voxelization strategy is applied on each face in a mesh to convert the irregular local structure into the regular volumetric representation. Second, a deep CNN is designed, which takes the volumetric representation as input, and outputs the learned filtered normals. All CNNs share the same workflow: three residual blocks, one max-pooling layer and four fully connected layers, of which the fourth layer outputs the filtered normals. At last, the vertex positions are updated according to the filtered normals. Moreover, we propose an iterative training framework for \textit{NormalNet}, in which the generation of training data and the training of CNNs are performed alternately, whereas the ground truth normals are taken as the guidance normals in GNF to get the target normals. Compared to the state-of-the-art schemes, \textit{NormalNet} can effectively generate more accurate filtering results and remove noise while preserving the original features and avoiding pseudo-features.


The rest of this paper is organized as follows. In the following section, we briefly summarize the related work. The proposed \emph{NormalNet} is introduced in Section 3. In Section 4, the training of \emph{NormalNet} is elaborated. Experimental results are presented in Section 5. Finally, Section 6 concludes the paper.

\section{Related work}
In this section, we briefly review the related work on the filtering-based mesh denoising and the neural-network-based 3D model processing.
\label{Voxelization}
\begin{figure*}
\centering
\includegraphics[width=0.95\linewidth]{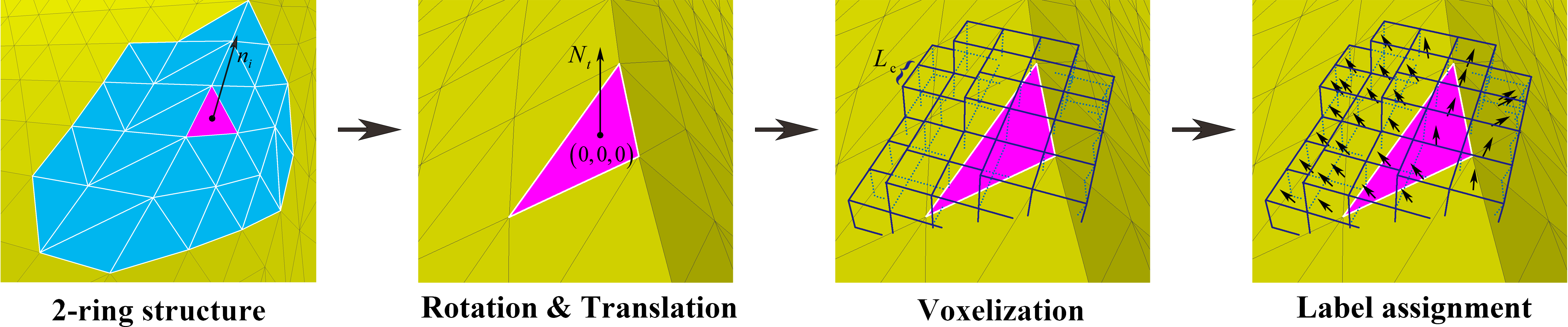}
\caption{Illustration of the proposed voxelization strategy. For a face in a mesh, a 2-ring patch is constructed. Two matrices that represent rotation and translation are computed for normalization. The irregular 3D structure around this face is split into small cubes. A label, which is the average normal of the faces in a cube, is then assigned to the cube.}
\label{fig_voxelization}
\end{figure*}
\subsection{Filtering-based mesh denoising}
Owing to the edge-preserving property of the bilateral filter, researchers have made many attempts to adopt the bilateral filtering in mesh denoising~\cite{fleishman2003bilateral,jones2003non,zheng2011bilateral}. Nevertheless, the photometric weights in the bilateral filter cannot be estimated accurately from a noise-corrupted mesh. The joint bilateral filter~\cite{Petschnigg}, in which the photometric weights are computed from a reliable guidance image, is proposed to improve the capability of bilateral filtering. Inspired by this idea, Zhang \textit{et al.}~\cite{zhang2015guided} propose the guided normal filtering, in which the guidance information is obtained as the average normal of a local patch. This scheme works well with respect to feature preservation but cannot achieve satisfactory results in regions with complex shapes and sometimes introduces pseudo-features. To overcome the problems of~\cite{zhang2015guided}, in the subsequent work~\cite{Li2017}, the guidance normals are computed by the corner-aware neighbourhood, which is adaptive to the shapes of corners and edges.

Recently, there have been increasing efforts to exploit the geometric attributes for mesh denoising. In~\cite{Zhang2015Variational}, the normal filtering is performed by means of a total variation, which assumes the normal change is piecewise constant. Wei \textit{et al.}~\cite{wei2015bi} propose to cluster faces into piecewise-smooth patches and refine face normals with the help of vertex normal fields. In~\cite{he2013mesh}, a differential edge operator is proposed and the L0 minimization is employed to remove noise while preserving the sharp features. Further more, Lu \textit{et al.}~\cite{LU2017} apply an additional vertex filtering before the L1-median face normal filtering, which proves to be capable of handling high noise levels and noise distributed in a random direction. However, feature information, such as edges and corners with less noise, may be blurred due to prefiltering. In~\cite{Yadav2017Robust}, the Tukey bi-weight similarity function is proposed to replace the similarity function in the computation of bilateral weights; in addition, an edge-weighted Laplace operator is introduced for vertex updating to reduce face normal flips. In~\cite{GGNF}, the graph-based feature detection is employed to construct accurate guidance normals; however, this method may introduce pseudo-features when the shape of the noise is complex, which is common in scanned models.

\subsection{Neural-Network-based 3D model processing}
Driven by the great success of deep learning in image processing, researchers in graphics are also attempting to employ deep neural networks for 3D model processing. However, due to the property of irregular connectivity, processing 3D models with neural networks remains challenging. Numerous works have focused on transforming 3D models into regular data. For instance, in~\cite{7410471,7273863}, 3D models are represented by 2D rendered images and panoramic views. Furthermore, some studies~\cite{WuSKYZTX15,7353481,abs-1712-01537,abs-1709-07599} have employed voxelization to transform models into regular 3D data. Moreover, in~\cite{abs-1709-04304,cgf.12844,YiSGG16,abs-1801-07829}, meshes are represented in the spectral or spatial domain for further processing.

In addition to these transform-based techniques, the direct application of neural networks to irregular data has also been extensively studied for point cloud data. PointNet~\cite{QiSMG16} is one of the first network architectures that can handle point cloud data. Subsequently, PointNet++~\cite{qi2017pointnet++} and the dynamic graph CNN~\cite{wang2018dynamic} are proposed to improve the network capability. Some attempts have been made to organize point clouds into structures. In~\cite{KlokovL17}, a kd-tree is constructed on a point cloud and is further used as the input of a neural network. A similar idea is presented in~\cite{RieglerUG16}, where the points are organized by an octree. Additional works~\cite{boulch2016deep,Rov18a,rakotosaona2019pointcleannet} focus on surface reconstruction, denoising and removing outliers.

Notably, in \cite{Wang-2016-SA}, Wang \textit{et al.} propose the filtered facet normal descriptor and model it with neural networks, however, these networks are not convolutional and only take face normal information into considered. In~\cite{abs-1809-05910}, the edge-based convolution and pooling operations are defined which can be directly on the constructs of the mesh.

\section{The Framework of NormalNet}

In this section, we introduce the framework of \emph{NormalNet}. Including four parts: the generation of patch, the introduction of GNF~\cite{zhang2015guided}, the voxelization strategy, and the proposed scheme.
\begin{figure*}
\centering
\includegraphics[width=0.95\linewidth]{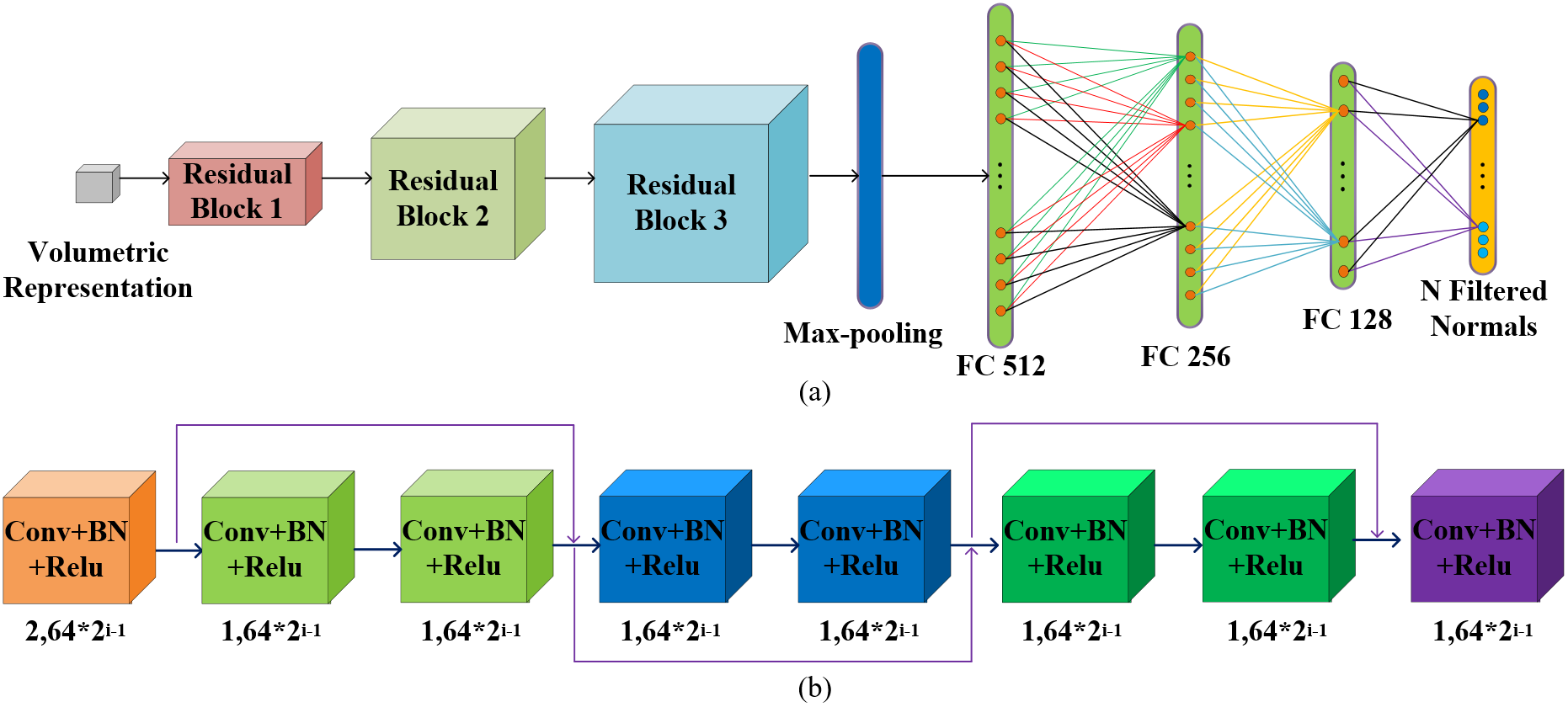}
\caption{ (a). The architecture of the deep network in \emph{NormalNet}. (b). The structure of residual blocks i,$[2,64*2^{i-1}]$ means that the convolution stride is 2 and the channel number is $64*2^{i-1}$.}
\label{fig3}
\end{figure*}
\subsection{The generation of patch}
\begin{figure}
\centering
\includegraphics[width=0.95\linewidth]{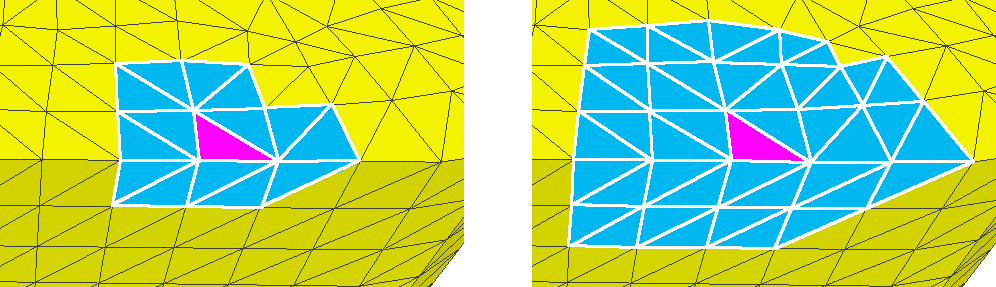}
\caption{Two examples of a 1-ring patch (left) and a 2-ring patch (right), $f_i$ is colored with purple.}
\label{fig33}
\end{figure}

As in mesh denoising, patch is a commonly used structure, so we describe the generation of $r$-ring patch first. Given a face $f_i$ as the center of patch $P$, an $r$-ring patch of $f_i$ is generated by finding all the faces that share at least one vertex with the faces in $P$, and adding them into $P$ for $r$ times. Two examples of 1-ring and 2-ring patches are shown in Fig.~\ref{fig33}.

\subsection{The guided normal filtering}
Since our scheme mimics the framework of GNF~\cite{zhang2015guided}, we briefly introduce GNF.

GNF is an iterative scheme, in which the face normal filtering is repeated for $N_f$ times. For a face $f_i$, the guided filtering is applied to obtain the denoised normal:
\begin{equation}
\n'_i = {e_i}\sum\limits_{{f_j} \in {{\mathcal{N}}_i}} {{a_j}{G_d}\left( {{c_i},{c_j}} \right){G_g}\left( {{\mathbf{g}_i},{\mathbf{g}_j}} \right){\mathbf{n}_j}}
\end{equation}
where $c_j$, $\n_j$ and $\mathbf{g}_j$ are the centre, face normal and guidance normal of $f_j$; ${\mathcal{N}_{i}}$ is a set of the geometrical neighbouring faces of ${f_i}$; ${e_i}$ is a normalization factor used to ensure that ${\n'_i}$ is a unit vector. ${G_d}$ and ${G_g}$ are the Gaussian kernels~\cite{test}, which are computed by:
\begin{equation}
{G_g} = \exp \left( { - \frac{{{{\left| {{\n_i}-{\n_j}} \right|}^2}}}{{2\mu_g^2}}} \right),
\end{equation}
\begin{equation}
{G_d} = \exp \left( { - \frac{{{{\left| {{c_i}-{c_j}} \right|}^2}}}{{2\mu_d^2}}} \right)
\end{equation}
where ${\mu_d}$ and ${\mu_g}$ are the Gaussian function parameters, ${\mu_d}$ is usually twice the average distance between adjacent face centres, ${\mu_g}$ is usually different for different meshes. Following the idea of~\cite{sun2007fast}, after each filtering, the position updating of the vertices is repeated for $N_v$ times to obtain the denoised mesh.

The guidance normal of $f_i$ is generated as follows. For each face suppose that $P$ is a 1-ring patch that contains $f_i$. The consistency $C\left( P \right)$ of $P$ is calculated as~\cite{zhang2015guided}:
\begin{equation}
\label{eq10}
C\left( P \right) = D\left( P \right)\cdot R\left( P \right)
\end{equation}
where $D\left( P \right)$ is the most significant face normal difference in $P$ and calculated by:
\begin{equation}
D\left( P \right) = \mathop {\max }\limits_{{f_i}, {f_j} \in P} \left| {{\mathbf{n}_i}-{\mathbf{n}_j}} \right|
\end{equation}
where $f_i$ and $f_j$ represent a pair of faces within $P$. $R\left( P \right)$ represents the saliency of $P$, which is computed by the saliency of edges in $P$:
\begin{equation}
R\left( P \right) = \frac{{\mathop {\max }\limits_{{e_i} \in P} \varphi \left( {{e_i}} \right)}}{{\varepsilon  + \sum\limits_{{e_i} \in P} {\varphi \left( {{e_i}} \right)} }}
\end{equation}
$\varepsilon$ is a small positive number to prevent the denominator from being zero, and $\varphi \left( {{e_i}} \right)$ is the saliency of an edge ${e_i}$:
\begin{equation}
\label{eq13}
\varphi \left( {{e_i}} \right) = \left| {{\mathbf{n}_{i1}} - {\mathbf{n}_{i2}}} \right|
\end{equation}
where ${\mathbf{n}_{i1}}$ and ${\mathbf{n}_{i2}}$ are the normals of the incident faces of ${e_i}$. Finally, the most consistent patch is chosen, and the average normal of the patch is regarded as the guidance normal.

The guidance normals generated by the above method have been proven to be effective on simple structures. However, the calculation of consistency is only based on the difference between face normals, and the structure information has not been fully considered. As mentioned before, scanned meshes may contain noise with huge face normal difference. This method will not be working well for such cases.

Rather than designing a method that works well on these meshes manually, we employed CNNs to obtain the learned filtered normals.

\subsection{The voxelization strategy}
\begin{figure*}
\centering
\includegraphics[width=0.95\linewidth]{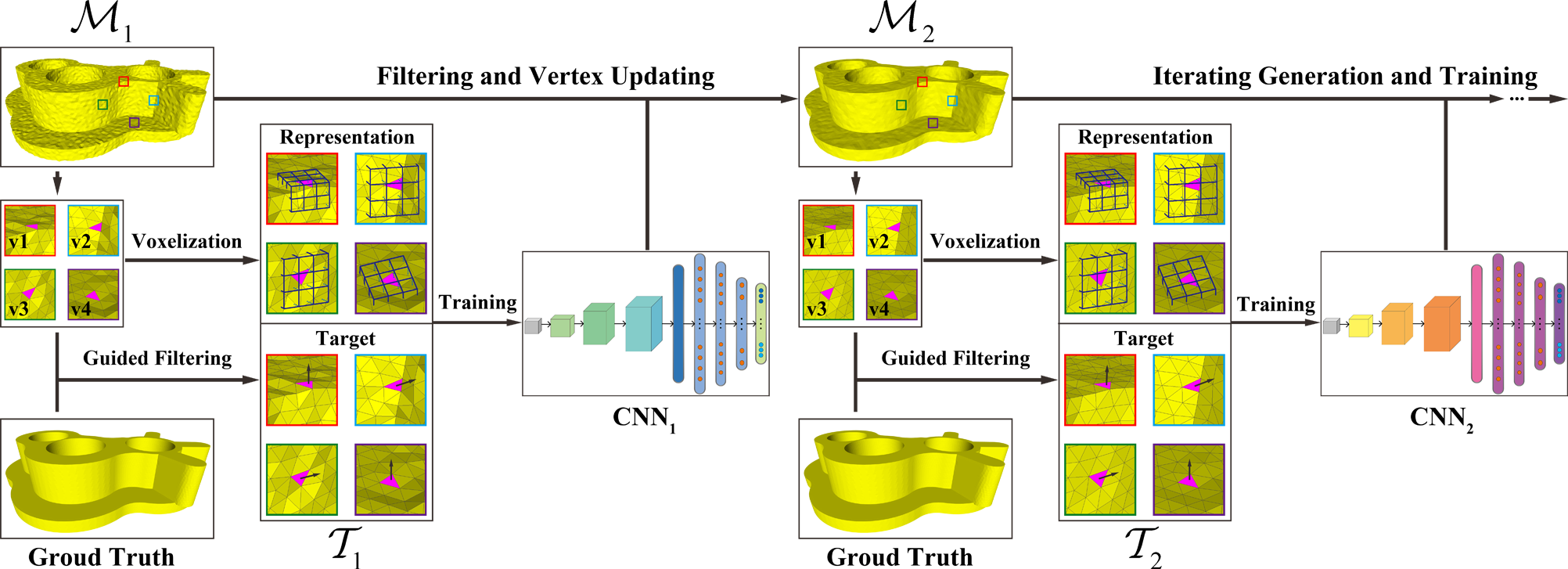}
\caption{The framework of generating training sets and training.}
\label{fig7}
\end{figure*}

The key to use CNNs for mesh denoising is the transformation of the irregular local structure around a face into a regular form such that the structure information is preserved and the CNN convolution operations are easily performed.

An illustration of the proposed voxelization strategy is shown in Fig.~\ref{fig_voxelization}. The normalization is applied to improve the robustness of the strategy first. The normalization process involves two operations: rotation and translation. In this way, all faces are normalized to a similar direction and position. Specifically, for a face $f_i$, a 2-ring patch is constructed. The average normal of this patch is $\n_i$. We then compute two matrices: $\W_r$, which represents the rotation from $n_i$ to a specific angle $N_t$, and $\W_t$, which represents the translation from the face centre $c_i$ to (0,0,0). The whole mesh is then rotated and translated by means of $\W_r$ and $\W_t$. Supposing $v_i$ is the coordinate of a vertex $i$ in the mesh, the new position of $v_i$ after normalization is:
\begin{equation}
{v'_i} = {\W_t}{\W_r}{v_i}
\end{equation}

After normalization, the space of the local mesh structure around $f_i$ is split into regular cubes denoted by $\left\{ {{B_{x,y,z}}\left| {x,y,z \in \left[ { - {T_s},{T_s}} \right]} \right.} \right\}$, where $T_s$ is the parameter that determines the number of cubes and  $B_{0,0,0}$ is located at the origin. The rest issue is to determine the size of each cube. In our work, the side length $L_c$ of the cubes is computed as:
\begin{equation}
{L_c} = \frac{d_s}{\alpha _c}
\end{equation}
where $d_s$ is the average distance between adjacent faces in the noisy mesh and ${\alpha _c}$ is the parameter that controls the size of the cubes.

For each cube, we employ the fast 3D triangle-box overlap testing strategy ~\cite{Akenine-Moller:2005:FTO:1198555.1198747} to find faces that overlap with this cube. If at least one face is overlaps with this cube, the label of this cube is assigned as the average normal of all the overlapped faces, denoted by $\mathcal{B}$; otherwise, the label is set to (0,0,0). In this way, we convert the irregular local mesh structure into the regular volumetric representation $\mathcal{V} = \left[\mathcal{B}\in\mathbb{R}^{{\left( {{{2T_s + 1}}} \right)}^3}\right]$. $\mathcal{V}$ is then used as the input of the network.

In our experiment, we set $T_s=20$, $\alpha _c = 8$ and $N_t = (0,1,0)$. Under these conditions, $\mathcal{V}$ is a 41x41x41x3 matrix that contains the most 3-ring structure around $f_i$. Each face is split into about 40$\sim$60 cubes, which is sufficient to represent the shape information. Smaller $T_s$ and $\alpha _c$ reduce the amount of information in $\mathcal{V}$ and lead to unsatisfactory results, whereas larger parameters can improve the performance slightly but greatly increase the training time.

\subsection{The proposed scheme}

The proposed scheme is also an iterative scheme which is repeated for $N_f$ times. During each iteration, for a face ${f_i}$ in a mesh, the voxelization strategy is employed to transform the irregular local mesh structure around ${f_i}$ into the regular volumetric representation. Then a CNN takes the volumetric representation as input, and outputs the filtered normals. Since the value of ${\mu_g}$ in GNF greatly affects the denoising results and is often different for different meshes. Therefore the output of the network contains $N$ filtered normals with different ${\mu_g}$. At last the positions of the vertices are updated according to the selected filtered normals by $N_v$ times.

The network architecture is shown in Fig.~\ref{fig3}. It contains three residual blocks, a global max-pooling layer and four fully connected layers. The numbers of channels of the residual blocks are 64, 128 and 256. All the convolution layers use $3*3*3$ filters except the first layer, which uses $5*5*5$ filters. Down-sampling is performed by a convolution operation with a stride of 2 in the first layer of each residual block. The network ends with four fully connected layers: the first three have 512, 256 and 128 channels. The fourth aims to predict the three coordinates of $N$ filtered normals and thus contains $3*N$ channels. All layers are equipped with batch normalization and ReLU, except the last layer is equipped with Tanh to ensure the output lies in [-1,1].

The network architecture is inspired by the philosophy of ResNet~\cite{7780459} and VGGNet~\cite{DBLP:journals/corr/SimonyanZ14a}. The purpose of \emph{NormalNet} is to estimate accurate filtered normals from the noisy signal. However, as the network goes deeper, abundant information beneficial to filtering normals from the input can vanish or "wash out" by the time it reaches the output layer. To address this problem, we adopt the shortcut connection from ResNet to directly pass the early feature map to the later layers. This greatly increases the forward flow of information and thus contributes to the prediction of face normals. In addition, during the backpropagation process, a shortcut path adds an extra component to the gradients compared to the plain network, which can mitigate the vanishing gradient problem, thereby accelerating the training process.

In our experiment, we set $N=6$, and the output of the CNN contains the filtering results of ${\mu_g=0.25, 0.3, 0.35, 0.4, 0.45, 0.5}$.

\section{NormalNet Training}

\begin{figure}
\begin{center}
\includegraphics[width=0.98\linewidth]{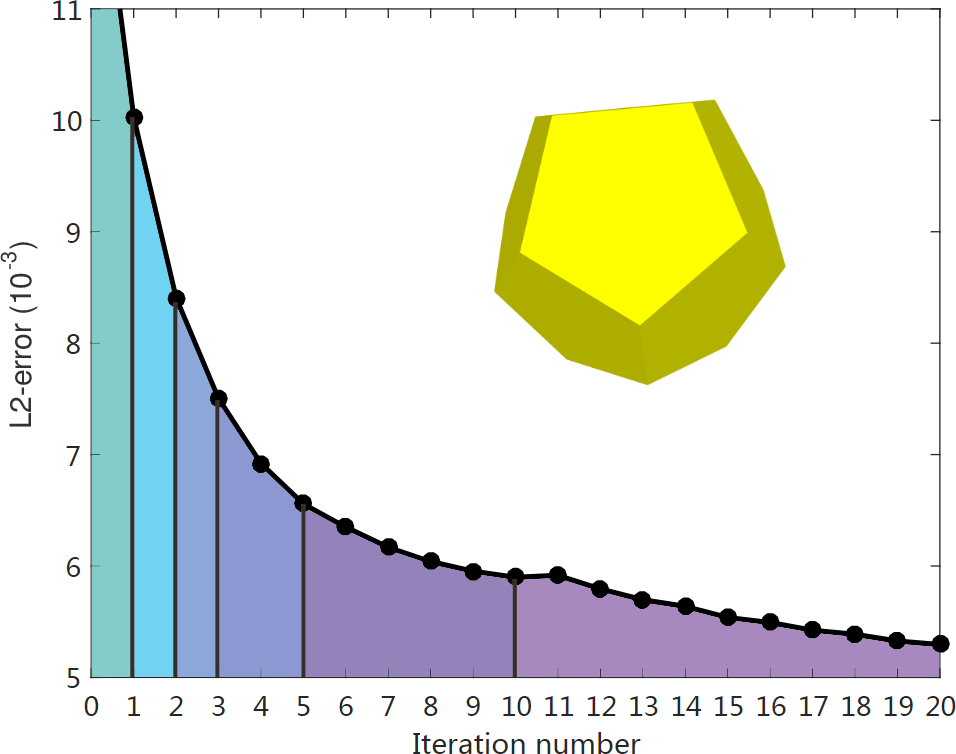}
\end{center}
\caption{Illustration of the L2 error results on the model \textit{Twelve}; each colour represents a CNN.}
\label{r9}
\end{figure}

\begin{table}
\begin{center}

\begin{tabular}{|c||c|c|c|c|c|c|}
\hline
 CNN$_i$ &    1 &         2 &     3 &      4 &     5  &     6 \\
\hline
Iteration numbers &         [1,1] &               [2,2] &        [3,3] &          [4,5]  &       [6,10]  &      [11,${N_f}$]   \\
\hline

\end{tabular}

\end{center}
\caption{The settings of the corresponding iteration numbers for each CNN$_i$.}
\label{table0}
\end{table}

\begin{table}

\begin{center}
\footnotesize{
\begin{tabular}{|c||c|c|c|c|c|}
\hline
 Para &    \emph{Fandisk} &         \emph{Table} &      \emph{Joint} &      \emph{Twelve} &     \emph{Block}   \\
\hline
\hline
        ${N_f}$ &         10 &               15 &         5 &          25  &       20  \\
\hline
        ${N_v}$ &         20 &                15 &        15 &          10  &        30  \\
\hline
        ${\mu_g}$  &       0.25 &           0.4 &        0.25 &       0.3  &      0.3  \\
\hline
\hline

 Para &    \emph{Bunny} &         \emph{Angel} &      \emph{Iron} &      \emph{Pierrot} &     \emph{Rocker-arm}   \\
\hline
\hline
        ${N_f}$ &         2 &               3 &         10 &          10  &       10  \\
\hline
        ${N_v}$ &         5 &                4 &        10 &          10  &        10  \\
\hline
        ${\mu_g}$  &       0.3 &           0.3 &        0.35 &       0.35  &      0.25  \\
\hline
\hline

 Para &    \emph{Eagle} &         \emph{Gargoyle} &      \emph{BallJoint} &      \emph{Boy01F} &     \emph{Boy02F}   \\
\hline
\hline
        ${N_f}$ &         4 &               5 &         4 &          14  &       7  \\
\hline
        ${N_v}$ &         5 &                10 &        10 &          20  &        20  \\
\hline
        ${\mu_g}$  &       0.4 &           0.3 &        0.4 &       0.4  &      0.35 \\
\hline
\hline

 Para &    \emph{Cone04V1} &         \emph{Girl02V1} &      \emph{Cone16V2} &      \emph{Girl01V2} &     -   \\
\hline
\hline
        ${N_f}$ &         20 &               15 &         10 &          3  &       -  \\
\hline
        ${N_v}$ &         20 &                20 &        10 &          15  &        -  \\
\hline
        ${\mu_g}$  &       0.45 &           0.45 &        0.3 &       0.4  &      -  \\
\hline

\end{tabular}
}
\end{center}
\caption{The settings of ${N_f}$, ${N_v}$  and ${\mu_g}$.}
\label{table_setting}
\end{table}
In this section, we introduce the training of \emph{NormalNet}. Including two parts: the iterative training and the training details.

\subsection{The iterative training}

The process of generating training sets and training is illustrated in Fig.~\ref{fig7}. For each CNN$_i$, a specified training data set $\mathcal{T}_i$ is generated from a group of meshes named by $\mathcal{M}_{i}$ and the corresponding ground truth. $\mathcal{T}_i$ is composed of numerous training tuples, each of which consists of a volumetric representation and $N$ target normals. For a face $f_i$ from a mesh in $\mathcal{M}_{i}$, the volumetric representation is obtained by applying the voxelization strategy on $f_i$. The target normals are obtained by employing GNF, whereas the ground truth normals are adopted as the guidance normals.
\begin{equation}
\n'_i = {e_i}\sum\limits_{{f_j} \in {{\mathcal{N}}_c}} {{a_j}{G_d}\left( {{c_i},{c_j}} \right){G_g}\left( {{\mathbf{gn}_i},{\mathbf{gn}_j}} \right){\mathbf{n}_j}}
\end{equation}
where $\mathbf{gn}_i$ and $\mathbf{gn}_j$ are the ground truth normals of $f_i$ and $f_j$, The other parameters are the same as defined in Eq.(1).

To make the training process balance with respect to various features. Suppose the maximum angle difference in the 2-ring patch of a face is $A_p$. All the faces in $\mathcal{M}_{i}$ are divided into 4 categories and we randomly select the same number of faces in each category for training:
\begin{itemize}
  \item v1: $\quad \quad \quad{A_p}>{80^\circ }$, large edge region.
  \item v2: ${50^\circ } < {A_p} \leq {80^\circ }$,  small edge region.
  \item v3: ${20^\circ } < {A_p} \leq {50^\circ }$,  curved region.
  \item v4: $ \quad \quad \quad{A_p} \leq {20^\circ }$, smooth region.
\end{itemize}

Initially, $\mathcal{M}_1$ is composed of noisy meshes that their ground truth are already known and without any processing. When $i>1$, $\mathcal{M}_i$ will be obtained by applying filtering on $\mathcal{M}_{i-1}$, which is performed by CNN$_{i-1}$, the parameters used in filtering are ${\mu_g=0.4}$ and $N_v=20$. The generation of the training data and the network training are alternately performed iteratively.

\subsection{Training details}

\begin{table*}
\begin{center}
\renewcommand{\arraystretch}{1.2}
\footnotesize
\begin{tabular}{|c|c|c||c|c|c|c|c|c|c|}
\hline
    model     &   Noise Level & $NormalNet$ &  Metrics   & L0M~\cite{he2013mesh}  & BI~\cite{wei2015bi}  & GNF~\cite{zhang2015guided}& CNR~\cite{Wang-2016-SA} &GGNF~\cite{GGNF}& \emph{NormalNet}  \\
\hline
\hline
\multirow{ 2}{*}{\emph{Fandisk}} & \multirow{ 2}{*}{0.3} & \multirow{ 2}{*}{$CS_1$} & ${E_v}\left( { \times {{10}^{{ - 3}}}}\right)$  &  1.850  & 1.509&   1.458  & 1.564  & 1.430 &  \textbf{1.281} \\
\cline{4-10}
 &   &  &  ${E_a}$  &    10.141  &  11.670  & 7.615    &  4.653   & 6.130   &\textbf{4.560} \\
\hline
\multirow{ 2}{*}{\emph{Table}} & \multirow{ 2}{*}{0.3} & \multirow{ 2}{*}{$CS_1$}  & ${E_v}\left( { \times {{10}^{{ - 3}}}} \right)$  &  1.961 &    1.571 &1.894 & 1.669& 1.378 & \textbf{1.372}  \\
\cline{4-10}
 &   &   &${E_a}$  &     \textbf{12.348}  &     18.635  &   17.544  & 18.912& 15.184  & 18.810 \\
\hline
\multirow{ 2}{*}{\emph{Joint}} & \multirow{ 2}{*}{0.2} & \multirow{ 2}{*}{$CS_1$} &   ${E_v}\left( { \times {{10}^{{ - 4}}}} \right)$  &  2.429   & 1.780&   1.428 & 1.434 & 1.438 &\textbf{1.403} \\
\cline{4-10}
 &  &   & ${E_a}$  &    11.181    & 5.489  &     5.920 &   4.390& 2.857  &  \textbf{2.625} \\
\hline
\multirow{ 2}{*}{\emph{Twelve}} & \multirow{ 2}{*}{0.5} & \multirow{ 2}{*}{$CS_1$}   & ${E_v}\left( { \times {{10}^{{ - 3}}}} \right)$ &  20.00   & 11.68&    5.955 &  7.427& 5.285 & \textbf{5.132} \\
\cline{4-10}
 &   &  &${E_a}$   &    12.147    & 20.038 &     11.099 & 5.734& 8.550 & \textbf{5.290}  \\
\hline
\multirow{2}{*}{\emph{Block}} & \multirow{ 2}{*}{0.4} & \multirow{ 2}{*}{$CS_1$}  &  ${E_v}\left( { \times {{10}^{{ - 3}}}} \right)$  &  9.273  & \textbf{4.895}&   5.417 &5.944 & 5.131 &5.331 \\
\cline{4-10}
 &   &  &${E_a}$   &    10.722    & 15.689 &     10.438  &   6.725  &10.007 &  \textbf{5.748}  \\
\hline
\multirow{2}{*}{\emph{Bunny}} & \multirow{ 2}{*}{0.2} & \multirow{ 2}{*}{$CS_1$}  &  ${E_v}\left( { \times {{10}^{{ - 6}}}} \right)$ &  7.897  & 7.727&   7.713&   7.879 &7.673 &\textbf{7.660}\\
\cline{4-10}
 &   &  &${E_a}$   &    9.359    & 11.008 &     7.494  &  7.649  &   7.246 &\textbf{6.963} \\
\hline
\multirow{ 2}{*}{\emph{Boy01F}} &\multirow{ 2}{*}{Scanned} & \multirow{ 2}{*}{$CS_1$}  &  ${E_v}\left( { \times {{10}^{{ - 4}}}}\right)$  &  8.119  & 8.120&   8.170  & 8.179& 8.106 & \textbf{8.098} \\
\cline{4-10}
 &   &  &${E_a}$   &    19.182  &       20.181  & 16.592 & 16.903& 16.266 &   \textbf{15.994}\\
\hline
\multirow{ 2}{*}{\emph{Boy02F}} & \multirow{ 2}{*}{Scanned} & \multirow{ 2}{*}{$CS_1$}  & ${E_v}\left( { \times {{10}^{{ - 3}}}}\right)$   &  8.446  & 8.514&   8.398  & 8.392& \textbf{8.302}  & 8.347  \\
\cline{4-10}
 &   &  &${E_a}$  &  16.601  &     17.056  & 14.491 & 14.931& 14.445  &   \textbf{13.966} \\
\hline
\multirow{ 2}{*}{\emph{Cone04V1}} &\multirow{ 2}{*}{Scanned} & \multirow{ 2}{*}{$CV_1$}  &  ${E_v}\left( { \times {{10}^{{ - 3}}}}\right)$  &  3.569  & 2.781&   2.657  & 2.806& 2.575 & \textbf{2.568} \\
\cline{4-10}
 &   &  &${E_a}$   &    37.618  &       22.144  & 15.658 & 15.670&  15.836&   \textbf{13.157}\\
\hline
\multirow{ 2}{*}{\textit{Girl02V1}}  & \multirow{ 2}{*}{Scanned} & \multirow{ 2}{*}{$CV_1$}  & ${E_v}\left( { \times {{10}^{{ - 3}}}}\right)$  & 1.934 & 1.899 & 1.751 & 1.658& 1.769& \textbf{1.634}  \\
\cline{4-10}
 &   &  &${E_a}$  & 37.826 & 26.353 & 19.707 & 20.121& 19.672& \textbf{17.903} \\
\hline

\multirow{ 2}{*}{\emph{Cone16V2}} &\multirow{ 2}{*}{Scanned} & \multirow{ 2}{*}{$CV_2$}  &  ${E_v}\left( { \times {{10}^{{ - 3}}}}\right)$  & 14.539 & 16.508 & 8.998 & 8.948&8.690 & \textbf{8.642} \\
\cline{4-10}
 &   &  &${E_a}$  & 27.872 & 12.642 & 10.805 & \textbf{8.468}& 9.862 & 8.731 \\
\hline
\multirow{ 2}{*}{\textit{Girl01V2}}  & \multirow{ 2}{*}{Scanned} & \multirow{ 2}{*}{$CV_2$}  & ${E_v}\left( { \times {{10}^{{ - 3}}}}\right)$ & 5.461 & 5.261 & 5.261 & 5.425& 5.226 & \textbf{5.171}  \\
\cline{4-10}
 &   &  &${E_a}$  & 28.401 & 18.098 & 18.098 & 14.627& 18.487 & \textbf{14.017} \\
\hline

\multirow{ 2}{*}{Average}  & \multirow{ 2}{*}{-} & \multirow{ 2}{*}{-}  & ${E_v}$ & 7.123 & 6.020 & 4.925 & 5.110& 4.750 & \textbf{4.719}  \\
\cline{4-10}
 &   &  &${E_a}$  & 19.449 & 16.583 & 12.955 & 11.565& 12.045 & \textbf{10.647} \\
\hline

\end{tabular}
\end{center}
\label{table1}
\caption{Performance comparisons between \emph{NormalNet} and the state-of-the-art methods.}
\end{table*}

The loss function is defined as the MSE between $N$ output normals and the target normals. We use the truncated normal distribution to initialize the weights and train the network from scratch. For the optimization method, we choose the Adam algorithm with a mini-batch size of 80, and the parameters for the Adam optimizer are $\beta_1=0.9$, $\beta_2=0.999$ and $\epsilon=1e-8$, which are the default settings in TensorFlow. The learning rate starts at 0.0001 and decays exponentially after 5000 training steps, for which the decay rate is 0.96. Each  CNN${_i}$ is trained individually. A test set that randomly selects some faces from test models is built for evaluation. The evaluation metric for the network is defined as the average angular error over the entire test set. Each network is trained for 10 epochs, and the average angular error is 1-3 degrees after 10 epochs. The network with the smallest error is selected for utilization.

In our experiment, we select 45000 faces in each category; thus, the total size of $\mathcal{T}_i$ is 180000. The training process is executed on a computer with an Intel Core i7-7700 CPU and NVIDIA GTX1080, and each epoch is approximately 3 hours. Increasing the number of channels, the number of layers or the size of $\mathcal{T}_i$ will not substantially improve the performance of the networks and only multiplies the training time. Halving the numbers of channels or the size of $\mathcal{T}_i$ will also halve the training time; however, the average angular error will increase to 4-6 degrees.

\section{Experimental Results}
In this section, the extensive experimental results are presented to demonstrate the performance of \emph{NormalNet}.

\subsection{Comparison study}
We perform the experimental comparisons on 19 test models, including 6 synthetic models: \textit{Joint}, \textit{Twelve}, \textit{Bunny}, \textit{Fandisk}, \textit{Table}, \textit{Block}; 4 scanned models collected from Internet where the type of scanner is unknown: \textit{Angel}, \textit{Iron}, \textit{Rocketarm}, \textit{Pierrot}; 6 scanned models which have rich features and generated by Microsoft Kinect v1, Microsoft Kinect v2 and Microsoft Kinect v1 via the Kinect-Fusion technique~\cite{Wang-2016-SA}, respectively: \textit{Core04V1}, \textit{Girl02V1}, \textit{Core16V2}, \textit{Girl01V2}, \textit{Boy01F}, \textit{Boy02F}; and 3 scanned models generated by laser scanners~\cite{8012522}: \emph{Eagle}, \emph{Gorgoyle} and \emph{BallJoint}. For the synthetic models, the noise type in \textit{Fandisk}, \textit{Table}, \textit{Bunny} and \textit{Block} is Gaussian white noise, while that of \textit{Joint} and \textit{Twelve} is impulsive noise.

We compare \emph{NormalNet} with several state-of-the-art algorithms in terms of objective and subjective evaluations. The compared algorithms are 1) guided normal filtering (GNF)~\cite{zhang2015guided}, 2) L0 minimization optimization (L0M)~\cite{he2013mesh}, 3) BI-normal filtering (BI)~\cite{wei2015bi}, 4) cascaded normal regression (CNR)~\cite{Wang-2016-SA}, 5) graph-based normal filtering (GGNF)~\cite{GGNF}, and 6) normal-voting-tensor-based scheme (VT)~\cite{8012522}. The source codes of GNF, L0M, BI, CNR and GGNF are kindly provided by their authors or implemented by a third party, while the author of VT provides the input models and their denoising results.

\subsection{Parameter settings}
As shown in Fig.~\ref{r9}, during the denoising process for most meshes, the L2-error decreases rapidly during the first three iterations and decreases slowly after ten iterations. In order to design a lightweight network, the iteration numbers are divided into six intervals, each of which corresponds to a specific CNN$_i$, as listed in Table~\ref{table0}. Thus, the training cost decreases by more than 70\% at the price of slightly decreased performance of CNN, the average angular error will increase 0.1-0.15 degree.

Three $NormalNet$, namely, $CV_1$, $CV_2$ and $CS_1$, are trained on Kinect-v1 training set (73 meshes), Kinect-v2 training set (73 meshes) and a remake of synthetic training set (60 meshes, where some meshes are excluded from the training sets for experiments) provided by~\cite{Wang-2016-SA}. The test models \emph{Cone04V1}, \emph{Girl02V1}, \emph{Cone16V2}, and \emph{Girl01V2} are denoised by the corresponding networks $CV_1$ and $CV_2$, and all the other test models are denoised by $CS_1$. The settings of the parameters $N_f$, $N_v$ and $\mu_g$ and the parameters in other schemes refer to the settings used in~\cite{zhang2015guided} and~\cite{GGNF}. The parameter settings of $N_f$, $N_v$ and $\mu_g$ are shown in Table~\ref{table_setting}.

\subsection{Objective performance comparison}
\begin{figure*}
\begin{center}
\includegraphics[width=0.98\linewidth]{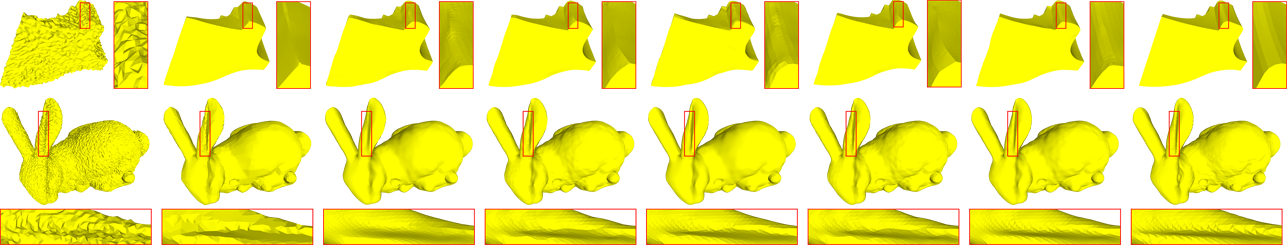}
{\footnotesize\\\hspace{0cm}(a)\hspace{1.95cm}(b)\hspace{1.95cm}(c)\hspace{1.95cm}(d)\hspace{1.95cm}(e)\hspace{1.95cm}(f)\hspace{1.95cm}(g)\hspace{1.95cm}(h)}
\end{center}
   \caption{Illustration of the denoising results on the models \textit{Fandisk} and \textit{Bunny}; the zoomed-in view of \textit{Bunny} has been rotated. (a) to (h) are the noisy mesh; the results of L0M~\cite{he2013mesh}, BI~\cite{wei2015bi}, GNF~\cite{zhang2015guided}, CNR~\cite{Wang-2016-SA}, GGNF~\cite{GGNF} and \emph{NormalNet}; and the ground truth.
}
\label{r1}
\end{figure*}

\begin{figure*}
\begin{center}
\includegraphics[width=0.98\linewidth]{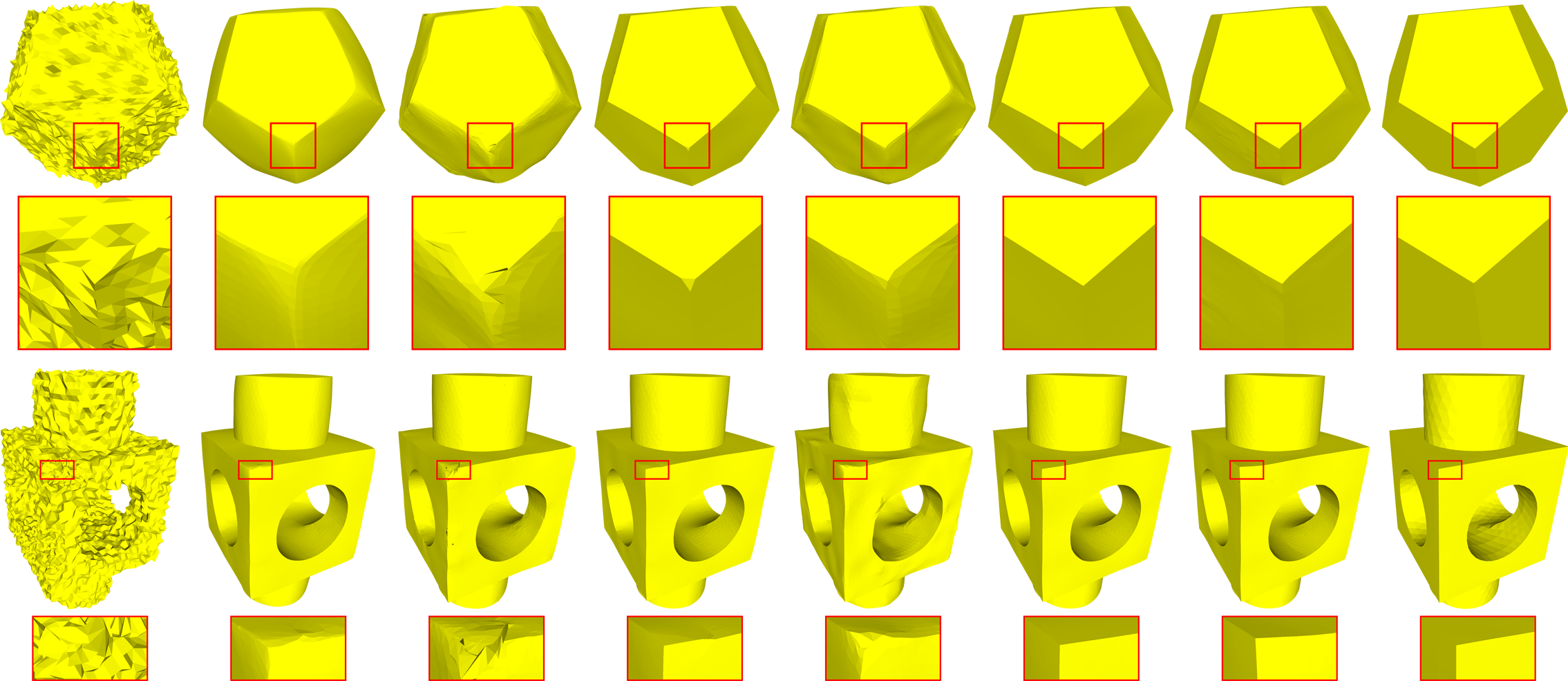}
{\footnotesize\\\hspace{0cm}(a)\hspace{1.95cm}(b)\hspace{1.95cm}(c)\hspace{1.95cm}(d)\hspace{1.95cm}(e)\hspace{1.95cm}(f)\hspace{1.95cm}(g)\hspace{1.95cm}(h)}
\end{center}
   \caption{Illustration of the denoising results on the models \textit{Twelve} and \textit{Block}. (a) to (h) are the noisy mesh; the results of L0M~\cite{he2013mesh}, BI~\cite{wei2015bi}, GNF~\cite{zhang2015guided}, CNR~\cite{Wang-2016-SA}, GGNF~\cite{GGNF} and \emph{NormalNet}; and the ground truth.
}
\label{r2}
\end{figure*}

Two error metrics~\cite{zheng2011bilateral} are employed to evaluate the objective denoising results of the models which have the ground truth:
\begin{itemize}
\item ${E_a}$: the mean angle square error, which represents the accuracy of the face normal;
\item ${E_v}$: the L2 vertex-based mesh-to-mesh error, which represents the accuracy of a vertex's position.
\end{itemize}

We compare the objective performance on 12 models. The comparison results of ${E_a}$ and ${E_v}$ are shown in Table 3, where the best results are bolded, \emph{NormalNet} performs best for 10 models on ${E_a}$ and 10 models on ${E_v}$, which achieves the best performance with respect to both metrics on most test models. CNR achieves the second best average results on ${E_a}$, which proves CNR is superior in estimating face normals. However, GNF and GGNF achieve better average results than CNR on ${E_v}$, which proves that filtering-based schemes perform better in recovering vertex positions. \emph{NormalNet} achieves the best average results on ${E_a}$ and ${E_v}$.

\subsection{Subjective Performance Comparison}

\subsubsection{Results on synthetic models}
\begin{figure*}
\begin{center}
\includegraphics[width=0.98\linewidth]{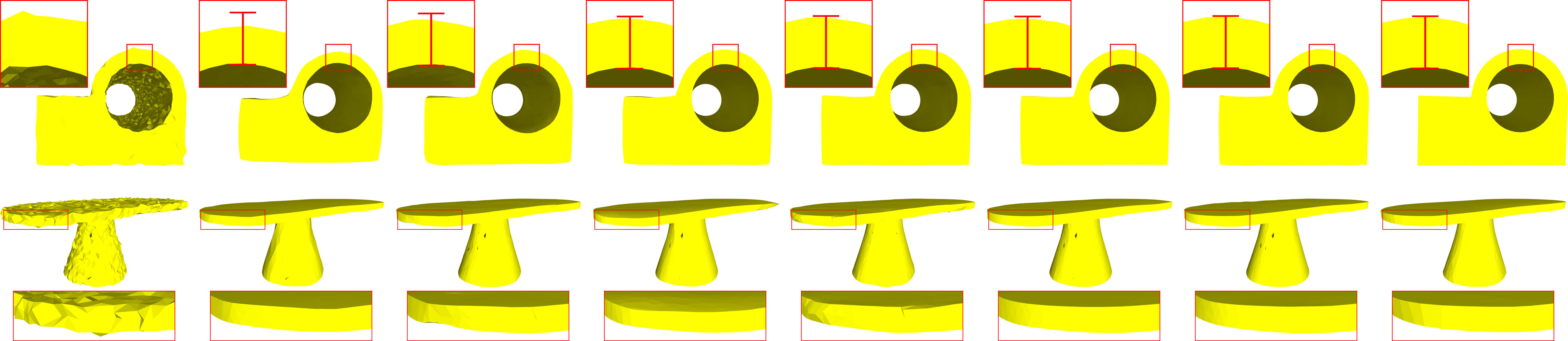}
{\footnotesize\\\hspace{0cm}(a)\hspace{1.95cm}(b)\hspace{1.95cm}(c)\hspace{1.95cm}(d)\hspace{1.95cm}(e)\hspace{1.95cm}(f)\hspace{1.95cm}(g)\hspace{1.95cm}(h)}
\end{center}
   \caption{Illustration of the denoising results on he tmodels \textit{Joint} and \textit{Table}. The red line in \textit{Joint} is the length of the ground truth. (a) to (h) are the noisy mesh; the results of L0M~\cite{he2013mesh}, BI~\cite{wei2015bi}, GNF~\cite{zhang2015guided}, CNR~\cite{Wang-2016-SA}, GGNF~\cite{GGNF} and \emph{NormalNet}; and the ground truth.
}
\label{r3}
\end{figure*}

The subjective performance comparison results of six synthetic models are illustrated in Figs.~\ref{r1}, ~\ref{r2} and~\ref{r3}.

Fig.~\ref{r1} presents the denoising results of two models with curved surfaces. The zoomed-in view illustrates that our scheme introduces fewer pseudo-features than other schemes. In Fig.~\ref{r2}, our scheme achieves similar performance to that of GGNF in these feature regions. The corner is recovered well, and the edge is sharp and clean. In \textit{Block}, the highlighted region in the red window has a higher triangulation density. Benefiting from the voxelization strategy, our scheme can preserve the structure information well and is thus less sensitive to the sampling irregularity. In Fig.~\ref{r3}, we perform a comparison on synthetic meshes with impulsive noise. In \textit{Table}, both our scheme and GGNF produce the best feature recovery results. In \textit{Joint}, the edge length of our scheme is closest to the ground truth.

\subsubsection{Results on scanned models}

\begin{figure*}
\begin{center}
\includegraphics[width=0.98\linewidth]{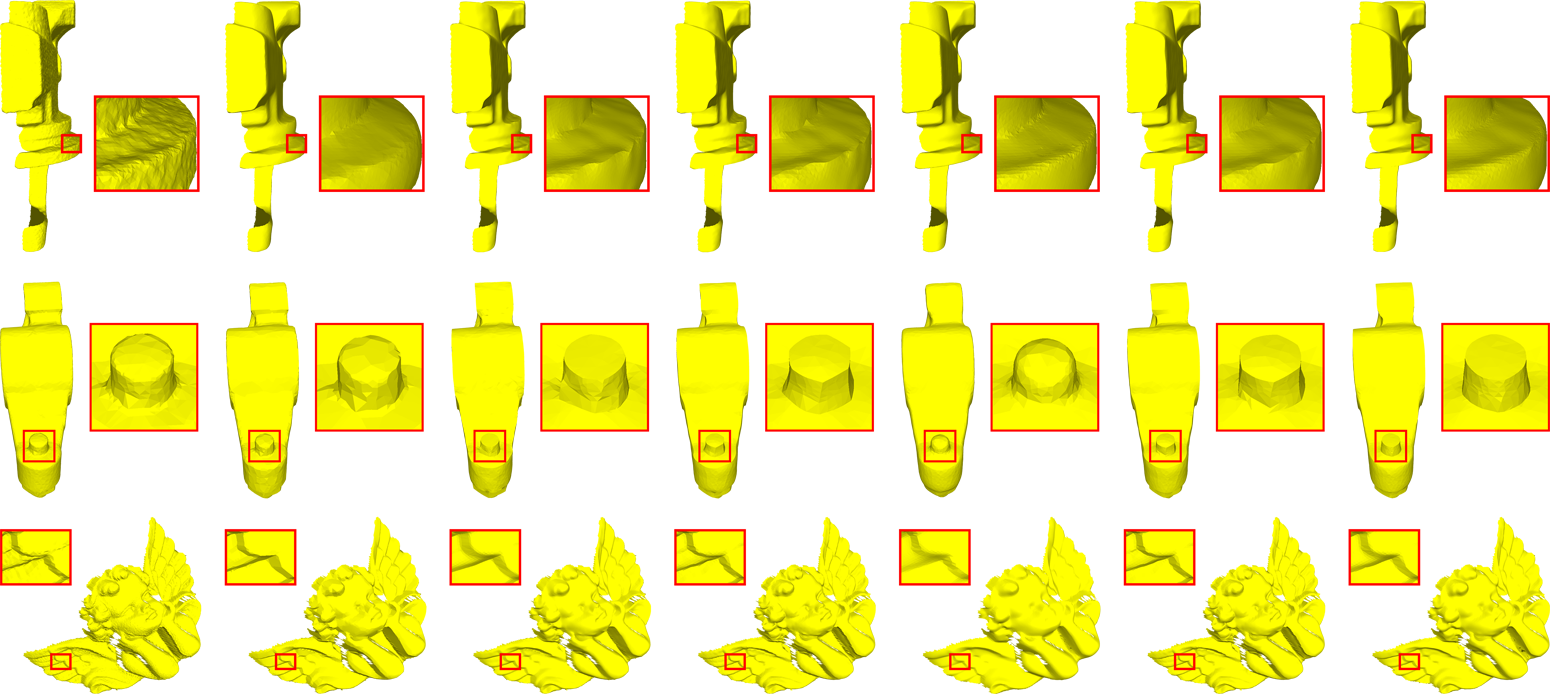}
{\footnotesize\\\hspace{0cm}(a)\hspace{2.2cm}(b)\hspace{2.2cm}(c)\hspace{2.2cm}(d)\hspace{2.2cm}(e)\hspace{2.2cm}(f)\hspace{2.2cm}(g)}
\end{center}
   \caption{Illustration of the denoising results on the models \textit{Iron}, \textit{Rocketarm} and \textit{Angel}, which are generated by unknown scanners. (a) to (g) are the noisy mesh and the results of L0M~\cite{he2013mesh}, BI~\cite{wei2015bi}, GNF~\cite{zhang2015guided}, CNR~\cite{Wang-2016-SA}, GGNF~\cite{GGNF} and \emph{NormalNet}.
}
\label{r4}
\end{figure*}

\begin{figure*}
\begin{center}
\includegraphics[width=0.98\linewidth]{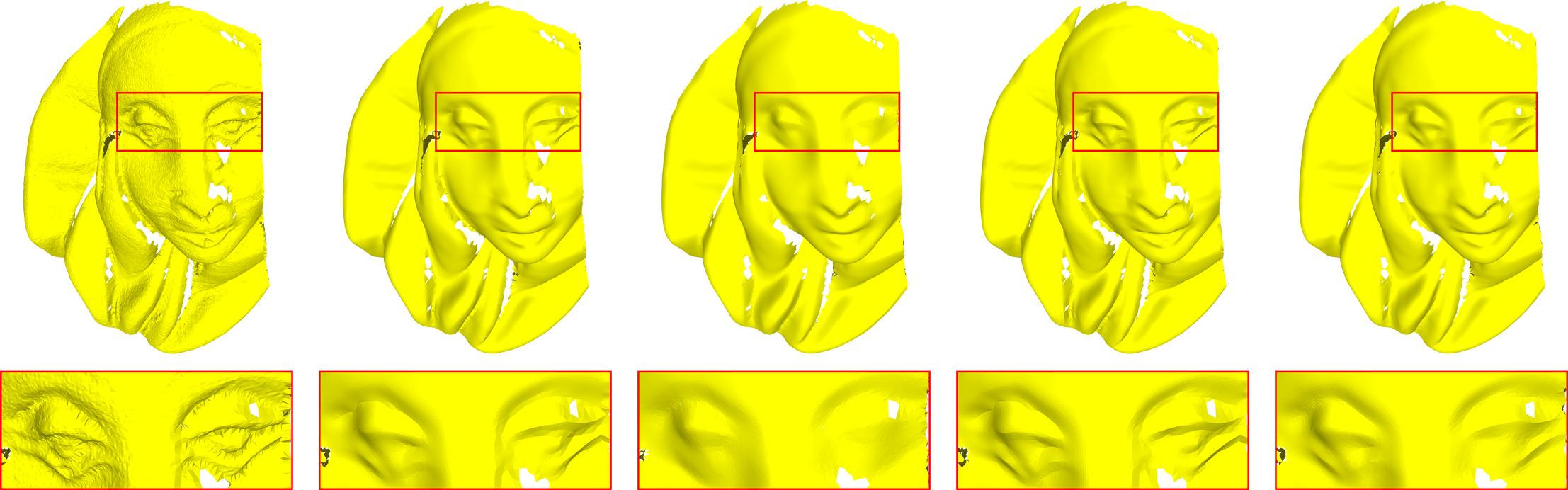}
{\footnotesize\\\hspace{0cm}(a)\hspace{3.3cm}(b)\hspace{3.3cm}(c)\hspace{3.3cm}(d)\hspace{3.3cm}(e)}
\end{center}
   \caption{Illustration of the denoising results on the scanned model \textit{Pierrot}, which is generated by unknown scanners. (a) to (e) are the noisy mesh and the results of GNF~\cite{zhang2015guided}, CNR~\cite{Wang-2016-SA}, GGNF~\cite{GGNF} and \emph{NormalNet}.
}
\label{r5}
\end{figure*}

We further provide the comparison results for models with different scanners. As illustrated in Figs.~\ref{r4} and~\ref{r5}, our scheme preforms well on models where the type of scanner is unknown. For fair comparison, CNR is also trained on the synthetic training set. In Fig.~\ref{r4}, for \textit{Iron}, both our scheme and CNR introduce fewer pseudo-features than the other schemes. However, in \textit{Rocketarm} and \textit{Angel}, CNR oversmooths the edges in the red boxes, whereas our scheme still produces satisfactory results. In Fig.~\ref{r5}, for \textit{Pierrot}, the region in the red box is corrupted by serrated noise. For GNF and GGNF, accurate guidance normals are difficult to compute under this type of noise. Thus, the denoising result is corrupted by pseudo-features. Furthermore, CNR succeeds in removing the serrated noise but fails to recover the edges around the eyes in the red box. Our scheme finds a balance between introducing pseudo-features and over-smoothing. The codes of L0M and BI could not process this region.

\begin{figure}
\begin{center}
\includegraphics[width=0.98\linewidth]{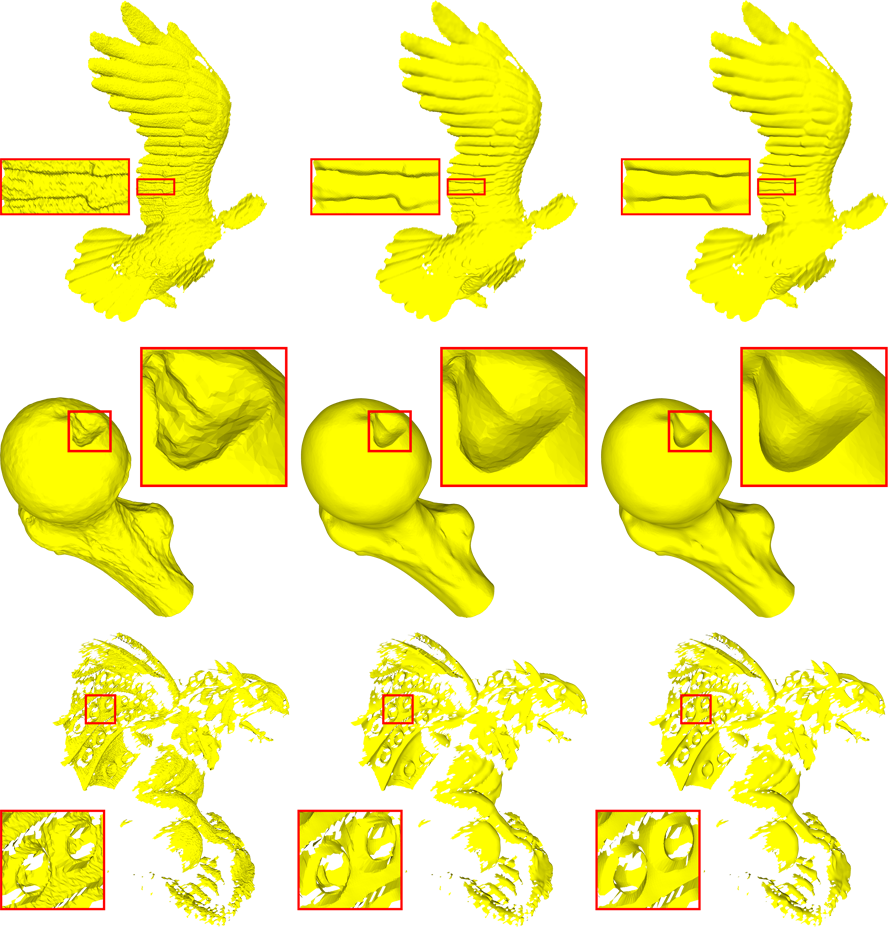}
{\footnotesize\\\hspace{0cm}(a)\hspace{2.7cm}(b)\hspace{2.7cm}(c)}
\end{center}
   \caption{Illustration of the denoising results of the scanned models \emph{Eagle}, \emph{Gargoyle} and \emph{BallJoint}, which are generated by laser scanners. (a) to (c) are the noisy mesh and the results of VT~\cite{zhang2015guided} and \emph{NormalNet}.
}
\label{r6}
\end{figure}

In Fig.~\ref{r6}, we compare \emph{NormalNet} with (VT)~\cite{8012522} on the models provided by the authors, which are generated by laser scanners. Our scheme produces better feature recovery results than VT on all three models that contain complex structures, which further verifies the capability of \emph{NormalNet}.

\begin{figure*}
\begin{center}
\includegraphics[width=0.98\linewidth]{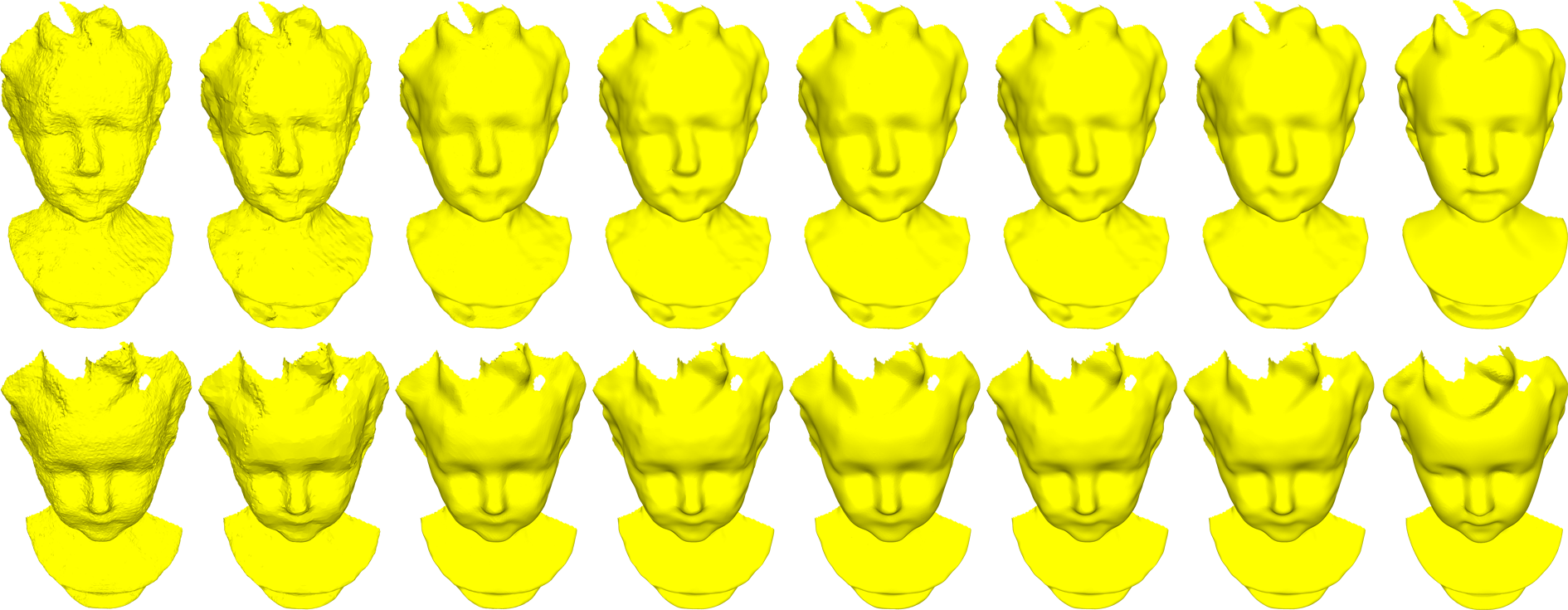}
{\footnotesize\\\hspace{0cm}(a)\hspace{1.95cm}(b)\hspace{1.95cm}(c)\hspace{1.95cm}(d)\hspace{1.95cm}(e)\hspace{1.95cm}(f)\hspace{1.95cm}(g)\hspace{1.95cm}(h)}
\end{center}
  \caption{Illustration of the denoising results on the models \textit{Boy01F} and \textit{Boy02F}, which are generated by Microsoft Kinect v1 via the Kinect-Fusion technique. (a) to (h) are the noisy mesh; the results of L0M~\cite{he2013mesh}, BI~\cite{wei2015bi}, GNF~\cite{zhang2015guided}, CNR~\cite{Wang-2016-SA}, GGNF~\cite{GGNF} and \emph{NormalNet}; and the ground truth.
}
\label{r7}
\end{figure*}

\begin{figure*}
\begin{center}
\includegraphics[width=0.98\linewidth]{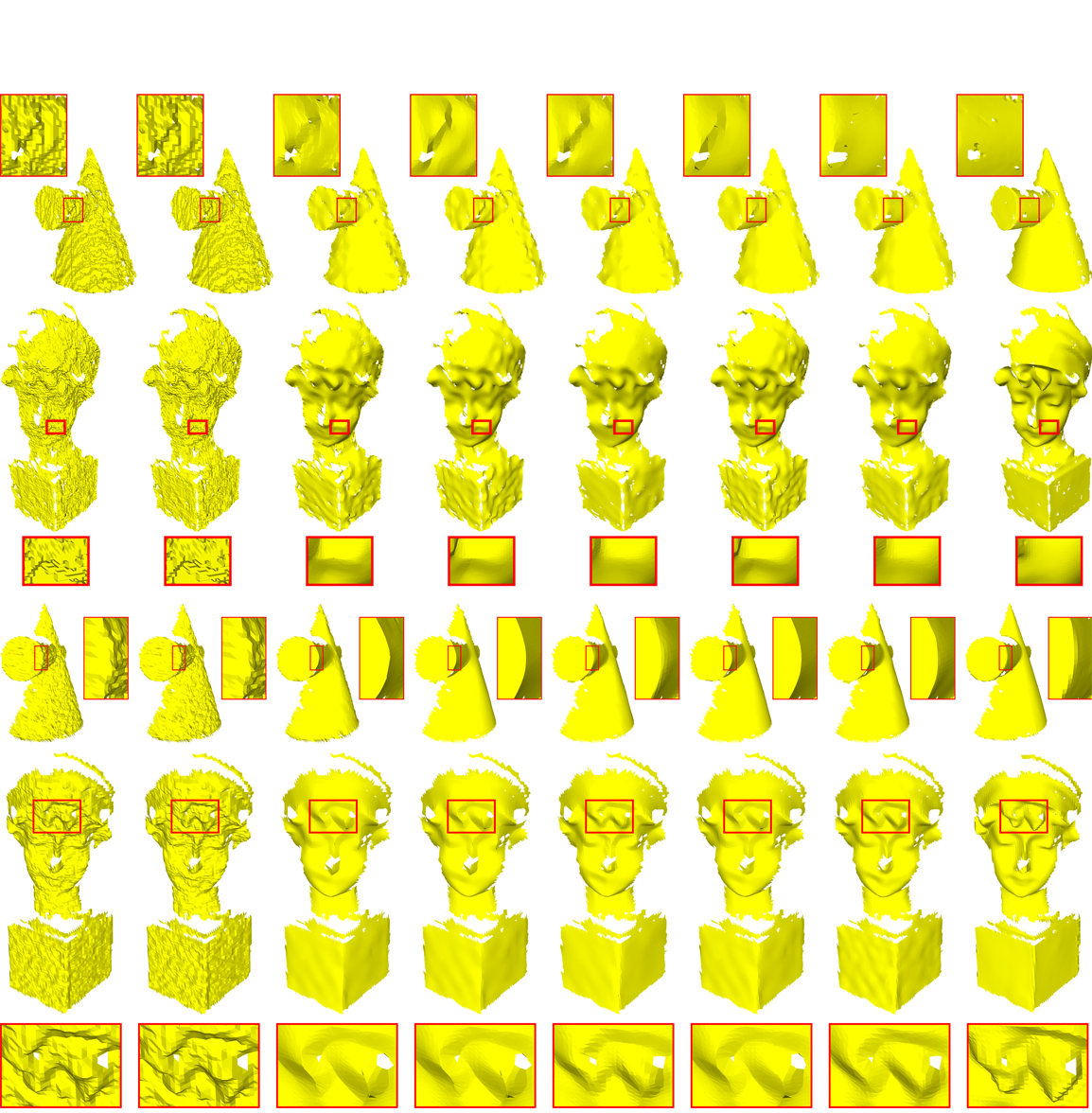}
{\footnotesize\\\hspace{0cm}(a)\hspace{1.95cm}(b)\hspace{1.95cm}(c)\hspace{1.95cm}(d)\hspace{1.95cm}(e)\hspace{1.95cm}(f)\hspace{1.95cm}(g)\hspace{1.95cm}(h)}
\end{center}
   \caption{Illustration of the denoising results on the models \textit{Cone04V1}, \textit{Girl02V1}, \textit{Cone16V2} and \textit{Girl01V2}, which are generated by Microsoft Kinect v1 and v2. (a) to (h) are the noisy mesh; the results of L0M~\cite{he2013mesh}, BI~\cite{wei2015bi}, GNF~\cite{zhang2015guided}, CNR~\cite{Wang-2016-SA}, GGNF~\cite{GGNF} and \emph{NormalNet}; and the ground truth.
}
\label{r8}
\end{figure*}

In Figs.~\ref{r7} and~\ref{r8}, the models are generated by Microsoft Kinect V1 and V2 and provided by the author of CNR. However, we do not have sufficient data to train \emph{NormalNet} for the models generated by Microsoft Kinect v1 via the Kinect-Fusion technique. Therefore, $CS_1$ is employed to denoise these models. In Fig.~\ref{r7}, our scheme outputs similar denoising results as CNR and GGNF. In Fig.~\ref{r8}, our scheme achieves the best smoothing result and the other schemes fail to remove noise in \textit{Cone04V1}. In \textit{Girl02V1} and \textit{Girl01V2}, both CNR and our scheme avoid introducing pseudo-features. In \textit{Cone16V2}, most schemes achieve similar feature recovery results.

\section{Conclusion}

In this paper, we present a learning-based normal filtering scheme for mesh denoising. The scheme maps the guided normal filtering into a deep network and follows the iterative framework of filtering-based scheme. During each iteration, first, to facilitate the 3D convolution operations, the voxelization strategy is applied on each face in a mesh to transform the irregular local structure into the regular volumetric representation. Second, instead of the guidance normal generation and the guided filtering in GNF, the output of voxelization is then input into a CNN to estimate accurate filtered normals. Finally, the vertex positions are updated according to the filtered normals. What's more, the iterative training framework is proposed for effectively training. The experimental results show that our scheme outperforms state-of-the-art works with respect to both objective and subjective quality metrics and can effectively remove noise while preserving the original features and avoiding pseudo-features.

\appendices

\ifCLASSOPTIONcaptionsoff
  \newpage
\fi

\bibliographystyle{IEEEtran}
\bibliography{IEEEabrv,egbib}

\end{document}